\documentclass{ws-p8-50x6-00}
\usepackage{latexsym,amssymb,amsfonts}

\newtheorem{theorem}{Theorem}

\begin{document}

\title{Gravitational waves and dynamics\\ 
of compact binary systems}

\author{Luc Blanchet}

\address{Institut d'Astrophysique de Paris (C.N.R.S.),\\ 
98$^{bis}$ boulevard Arago, 75014 Paris, France\\ 
E-mail: blanchet@iap.fr}

\maketitle

\abstracts{Part A of this article is devoted to the general
investigation of the gravitational-wave emission by post-Newtonian
sources. We show how the radiation field far from the source, as well
as its near-zone inner gravitational field, can (in principle) be
calculated in terms of the matter stress-energy tensor up to any order
in the post-Newtonian expansion. Part B presents some recent
applications to the problems of the dynamics and gravitational-wave
flux of compact binary systems. The precision reached in these
developments corresponds to the third post-Newtonian approximation.}

\section{Introduction}\label{I}

The theory of gravitational radiation from isolated sources, in the
context of general relativity, takes on importance because of the
possibility of comparing the theory with contemporary astrophysical
observations. We have in mind the large-scale optical interferometers
LIGO, VIRGO, GEO and TAMA, which should routinely observe the
gravitational waves produced by massive and rapidly evolving systems
such as inspiralling compact binaries.  To prepare these experiments
the required theoretical work consists of (i) carrying out a
sufficiently general solution of the Einstein field equations, valid
for a large class of matter systems, (ii) applying the latter solution
to specific systems like binaries.

For general isolated sources the basic problem is to relate the
asymptotic gravitational-wave form, at the location of a detector in
the source's wave zone, to the stress-energy tensor $T^{\alpha\beta}$
of the matter fields. In this article we shall explore this problem
using the post-Newtonian expansion, when the speed of light $c\to
+\infty$, as the basic approximation. This approximation has proved in
the past to constitute an invaluable tool for understanding many
aspects of the problems of radiation and motion in general relativity
(for reviews see
Refs. \cite{D300,Th300,W94,Bhouches,Behlers,Bliving}).

Concerning the application to compact binaries, the point is that a
model made of two structureless point-particles, characterized by two
mass parameters $m_1$ and $m_2$ (and possibly two spins), should be
sufficient to describe the ``inspiral'' phase, driven by gravitational
radiation reaction, which precedes the final binary coalescence.  Most
of the non-gravitational effects, usually plaguing the dynamics of
binary star systems, are dominated by (post-Newtonian) gravitational
effects. However the real justification of the model of point
particles is that the effects due to the finite size of the compact
bodies are small \footnote{For instance the influence of the Newtonian
quadrupole moments, $Q_1$ and $Q_2$, induced by tidal interaction
between two compact objects (for which $a_1\simeq \frac{Gm_1}{c^2}$
and $a_2\simeq \frac{Gm_2}{c^2}$), is, by a simple Newtonian argument,
comparable (formally) with gravitational effects at the 5PN
order.}. Inspiralling compact binaries are ideally suited for
application of a post-Newtonian wave generation formalism, because it
has been shown \cite{3mn,CFPS93} that the post-Newtonian precision
required to implement successively the optimal filtering technique in
the LIGO/VIRGO detectors corresponds grossly, in the case of
neutron-star binaries, to the 3PN approximation, namely $1/c^6$ beyond
the quadrupole moment formalism.

In this article we review (Part A) and apply (Part B) a
wave-generation formalism based on systematic multipole decompositions
of Einstein's theory, and initiated by Blanchet and Damour
\cite{BD86,B87}, following preceding work by Bonnor \cite{Bo59} and
Thorne \cite{Th80}. The basic multipole moments in this approach are
some appropriately defined {\it source} moments, which are some
functions of time, supposed to describe the actual physical
source. They are iterated by means of a post-Minkowskian expansion of
the vacuum field equations (valid in the source's
exterior). Technically, the post-Minkowskian approximation scheme is
greatly simplified by the assumption of multipolar expansion, because
one can consider separately the iteration of the different multipole
pieces composing the exterior field. In our formalism, the radiation
field at future null infinity is described by the standard {\it
radiative} multipole moments \cite{Th80}, which are obtained as some
explicit non-linear functionals of the more basic source moments
\cite{BD92,B98quad,B98tail}. In the current situation, the {\it
closed-form} expressions of the source multipole moments have been
established in the case where the source is post-Newtonian
\cite{B95,B98mult} (existence of a small parameter which can be
regarded as a slow motion estimate $\sim v/c$). For post-Newtonian
sources the domain of validity of the post-Newtonian expansion ({\it
viz} the near zone) overlaps the exterior weak-field region, so that
there exists an intermediate zone in which the post-Newtonian and
multipolar expansions can be matched together \footnote{This is an
application of the method of matched asymptotic expansions
\cite{BuTh70,Bu71}.}. Furthermore, the complete expression of the
gravitational field inside the source, valid formally up to any
post-Newtonian order, has also been determined by the latter matching
\cite{PB01}.

The post-Newtonian wave-generation formalism permitted to compute the
total energy flux, ${\cal L}$, emitted by inspiralling compact
binaries. ${\cal L}$ was completed first to the 2PN order by Blanchet,
Damour and Iyer \cite{BDI95}, and, independently, by Will and Wiseman
using their own formalism \cite{WWi96,PW00}. Higher-order tail effects
at the 2.5PN and 3.5PN orders, as well as a crucial contribution of
tails generated by the tails themselves (the so-called ``tails of
tails'') arising at the 3PN order, were subsequently obtained
\cite{B98tail}. Notice that unlike the 1.5PN, 2.5PN and 3.5PN orders
which are entirely composed of tail terms, the 3PN order involves
also, besides the tails of tails, many non-tail contributions coming
from the relativistic corrections in the binary's multipole
moments. These have been recently computed by Blanchet, Iyer and
Joguet \cite{BIJ01}.

The dynamics of compact binaries, i.e. their equations of motion, is
also needed in this theoretical work. The 1PN approximation in the
equations of motion is standard \cite{EIH}. The 2PN terms have been
derived by Damour and Deruelle \cite{DD81a,D83a}, Kopeikin and
Grishchuk \cite{Kop85,GKop86}, Blanchet, Faye and Ponsot \cite{BFP98},
Itoh {\it et al} \cite{IFA01}, Pati and Will \cite{PW02}.  The 3PN
order has been obtained independently by two groups~: Jaranowski and
Sch\"afer \cite{JaraS98,JaraS99}, Damour, Jaranowski and Sch\"afer
\cite{DJS00,DJS01b} employ the ADM-Hamiltonian formalism of general
relativity; Blanchet and Faye \cite{BF00,BFeom,BFreg,BFregM}, de
Andrade, Blanchet and Faye \cite{ABF01} compute directly the equations
of motion (instead of a Hamiltonian) in harmonic coordinates. To the
3PN order the end results have been shown to be physically equivalent
in the sense that there exists a unique ``contact'' transformation of
the binary's dynamical variables, that changes the
harmonic-coordinates Lagrangian of Ref. \cite{ABF01} into a new
Lagrangian whose Legendre transform coincides exactly with the
Hamiltonian given in Ref. \cite{DJS00}. This equivalence is of course
quite satisfying.

\bigskip
\centerline{\bf{PART A~: POST-NEWTONIAN SOURCES}}
\bigskip

We describe the general solution of the field equations, valid for
isolated post-Newtonian sources, in a series of mathematical
statements, presented here without proofs. This presentation has the
advantage of emphasizing some key results and outlining the basic
``skeleton'' of the formalism. We define $h^{\alpha\beta} =
\sqrt{-g}\, g^{\alpha\beta} - \eta^{\alpha\beta}$, where
$g^{\alpha\beta}$ is the metric, $\eta^{\alpha\beta}$ is the Minkowski
metric, and $g = {\rm det}( g_{\alpha\beta})$. The field equations in
harmonic coordinates, $\partial_\beta h^{\alpha\beta} = 0$, read

\begin{eqnarray}
\Box h^{\alpha\beta} &=& \frac{16\pi
G}{c^4}\tau^{\alpha\beta}\;,\label{1}\\
\tau^{\alpha\beta}&\equiv&|g|T^{\alpha\beta}+\frac{c^4}{16\pi
G}\Lambda^{\alpha\beta}(h,\partial h,\partial^2h)\;,\label{2}
\end{eqnarray}
where $\Box$ is the flat d'Alembertian operator, $T^{\alpha\beta}$ is
the matter tensor and $\Lambda^{\alpha\beta}$ denotes the non-linear
gravitational source term.

\section{Linearized vacuum equations~: general solution}\label{II}

The first result concerns the solution of the linearized field
equations in the vacuum region exterior to the isolated source.

\begin{theorem}\label{th1}\cite{Th80,BD86} The most general solution of 
the linearized vacuum field equations, $\Box h^{\alpha\beta}_1 = 0 =
\partial_\beta h^{\alpha\beta}_1$, outside some time-like world tube
enclosing the source, and stationary in the past (when $t\leq -{\cal T}$), reads

\begin{equation}\label{3}
h^{\alpha\beta}_1 = k^{\alpha\beta}_1 + \partial^\alpha\varphi^\beta_1 +
\partial^\beta\varphi^\alpha_1 -
\eta^{\alpha\beta}\partial_\gamma\varphi^\gamma_1\;.
\end{equation}
The vector $\varphi^\alpha_1$ parametrizes a linearized gauge
transformation. The first term is given by \footnote{Multi-indices are
denoted by $L=i_1i_2\cdots i_l$; the derivative operator $\partial_L$ is
a short-hand for $\partial_{i_1}\cdots\partial_{i_l}$; a tensor $I_L$
is STF if and only if for any pair of indices $i_p$, $i_q\in L$, we
have $I_{\cdots i_p\cdots i_q\cdots}=I_{\cdots i_q\cdots i_p\cdots}$
and $\delta_{i_pi_q}I_{\cdots i_p\cdots i_q\cdots}=0$. The superscript
$(p)$ refers to the time derivation.}

\begin{eqnarray}
k^{00}_1 &=& -\frac{4}{c^2}\sum_{l\geq 0} \frac{(-)^l}{l !} \partial_L
 \!\!\left( \frac{1}{r} I_L (u)\right)\;,\label{4}\\ k^{0i}_1 &=&
 \frac{4}{c^3}\sum_{l\geq 1} \frac{(-)^l}{l !} \left\{ \partial_{L-1}
 \!\!\left( \frac{1}{r} I_{iL-1}^{(1)} (u)\right) + \frac{l}{l+1}
 \varepsilon_{iab} \partial_{aL-1} \!\!\left( \frac{1}{r} J_{bL-1}
 (u)\right)\right\}\;,\label{5}\\ k^{ij}_1
 &=&-\frac{4}{c^4}\sum_{l\geq 2} \frac{(-)^l}{l !} \left\{
 \partial_{L-2} \!\!\left( \frac{1}{r} I_{ijL-2}^{(2)} (u)\right) +
 \frac{2l}{l+1} \partial_{aL-2} \!\!\left( \frac{1}{r}
 \varepsilon_{ab(i} J_{j)bL-2}^{(1)}
 (u)\right)\right\}\;,\nonumber\\\label{6}
\end{eqnarray}
where $I_L$ and $J_L$ are two sets of ${\rm STF}$-tensorial multipole
moments, functions of the retarded time $u=t-r/c$, and satisfying the
conservation laws~: $I = {\rm const}$, $I_i = {\rm const}$, $J_i =
{\rm const}$.
\end{theorem}

\section{Exterior vacuum solution~: post-Minkowskian iteration}\label{III}

Consider the so-called post-Minkowskian expansion, or formal series in
powers of Newton's constant $G$,

\begin{equation}\label{7}
h^{\alpha\beta}_{\rm ext}=\sum_{n=1}^{+\infty}
G^nh_n^{\alpha\beta}\;,
\end{equation}
that is based on the general linearized exterior solution
$h_1^{\alpha\beta}[I_L,J_L]$ of Theorem \ref{th1}.  We obtain a
particular solution, at any post-Minkowskian order $n$, in the form
\cite{BD86}

\begin{equation}\label{8}    
h_n^{\alpha\beta} = {\cal FP}\, \Box^{-1}_{\rm ret}
[\Lambda_n^{\alpha\beta}] +v_n^{\alpha\beta}\;,
\end{equation}
where $\Box^{-1}_{\rm ret}$ is the usual retarded integral and ${\cal
FP}$ a specific operation (defined by complex analytic continuation)
of taking the finite part. The finite part is necessary because we are
looking for the exterior solution, in the form of a multipole
expansion which is singular at the origin $r=0$, supposed to be
located inside the compact-support source. The term
$v_n^{\alpha\beta}$ is a particular source-free retarded solution
constructed in such a way that the harmonic coordinate condition
($\partial_\beta h_n^{\alpha\beta}=0$) is satisfied.

\begin{theorem}\label{th2}\cite{BD86}
The most general solution of the Einstein field equations, in the
vacuum region outside the source, admitting post-Minkowskian and
multipolar expansions (and stationary in the past), is given by the
construction (\ref{7})-(\ref{8}). It depends on the source multipole moments
$I_L(u)$ and $J_L(u)$, and on four supplementary functions
parametrizing the gauge vector in Eq. (\ref{3}).
\end{theorem}

\section{General structure of the exterior solution}\label{IV}

The solution is valid in all the (weak-field) exterior region of the
source. We investigate its expansion when $r\to 0$ and $r\to +\infty$.

\begin{theorem}\label{th3}\cite{BD86}
The structure of the expansion of the post-Minkowskian exterior metric
in the near-zone (when $r\to 0$) is of the type \footnote{We denote by
${\hat n}_L$ (or sometimes $n_{<L>}$) the STF projection of the
product of unit vectors $n_L=n_{i_1}\cdots n_{i_l}$; an expansion into
STF tensors ${\hat n}_L={\hat n}_L(\theta,\phi)$ is equivalent to the
usual expansion in spherical harmonics
$Y_{lm}(\theta,\phi)$. Similarly we pose $x_L=x_{i_1}\cdots
x_{i_l}=r^l n_{L}$ and ${\hat x}_L=r^l{\hat n}_L$.}~: $\forall N \in
\mathbb{N}$,

\begin{equation}\label{9}
 h_n^{\alpha\beta}({\bf x},t) = \sum {\hat n}_L r^m (\ln r)^p
F^{\alpha\beta}_{L,m,p,n}(t)+o(r^{N}) \;,
\end{equation} 
where $m\in \mathbb{Z}$, with $m_0 \leq m \leq N$ (and $m_0$ becoming
more and more negative when $n$ grows), $p \in \mathbb{N}$ with $p \leq
n-1$. The functions $F^{\alpha\beta}_{L,m,p,n}$ are multilinear functionals of the
source multipole moments.
\end{theorem}
This theorem shows in particular that the general form of the
post-Newtonian expansion (i.e. when $c\to +\infty$) is necessarily of
the type $\sum (\ln c)^p/c^m$.

\begin{theorem}\label{th4}\cite{B87}
The most general multipolar and post-Minkowskian solution, stationary
in the past, (i) admits radiative coordinates $(T,{\bf X})$, for which
the expansion at infinity, $R\to +\infty$ with $U\equiv T-R/c= {\rm
const}$, takes the form~:

\begin{equation}\label{10}
H^{\alpha\beta}_n({\bf X},T) = \sum \frac{{\hat N}_L}{R^k}
K^{\alpha\beta}_{L,k,n}(U)+{\cal O}\left(\frac{1}{R^N}\right) \;,
\end{equation} 
where the functions $K^{\alpha\beta}_{L,k,n}$ are computable
functionals of the source multipole moments, (ii) is asymptotically
simple in the sense of Penrose \cite{P65}, perturbatively to any
post-Minkowskian order.
\end{theorem}
In radiative coordinates the retarded time $U$ is a null coordinate at
future null infinity. Theorem \ref{th4} permits to connect our work
with the study by Bondi {\it et al} \cite{BBM62} of the asymptotic
structure of radiative gravitational fields. Most importantly, it
permits to calculate the relations between the source multipole
moments $I_L$ and $J_L$ and the radiative ones, say $U_L$ and $V_L$,
which parametrize the asymptotic wave form. These relations involve
many non-linear multipole interactions associated with tails and
related effects \cite{BD92,B98quad,B98tail}.

\section{Post-Newtonian wave generation}\label{V}

Suppose that there exists a solution $h^{\alpha\beta}$, valid
everywhere inside and outside a post-Newtonian source, whose multipole
expansion coincides with the previously constructed exterior field~:
${\cal M}(h^{\alpha\beta})\equiv h^{\alpha\beta}_{\rm ext}$. Our basic
assumption is that of the matching between the near-zone expansion,
$r\to 0$, of the exterior multipole expansion~: i.e. $\overline{{\cal
M}(h^{\alpha\beta})}$, and the far-zone expansion, $r\to +\infty$, of
the inner post-Newtonian expansion~: i.e. ${\cal
M}(\overline{h^{\alpha\beta}})$. The notation is ${\cal M}$ for the
operation of taking the multipole expansion, and an overline for the
post-Newtonian (re-)expansion. As shown by the matching, the near-zone
expansion of the outer field is actually a post-Newtonian expansion,
and the far-zone expansion of the inner field is identical to the
multipole decomposition. The matching equation reads

\begin{equation}\label{11}
\overline{{\cal M}(h^{\alpha\beta})}=\sum {\hat n}_L r^m (\ln r)^p
F^{\alpha\beta}_{L,m,p}(t)={\cal M}(\overline{h^{\alpha\beta}})\;,
\end{equation}
where we have given the common structure of both types of asymptotic
expansions, which is given, after formal post-Minkowskian
re-summation, by Theorem \ref{th3}.

\begin{theorem}\label{th5}\cite{B95,B98mult}
Under the hypothesis of matching, Eq. (\ref{11}), the multipole
expansion of the field exterior to a post-Newtonian source reads 

\begin{equation}\label{12}    
{\cal M}(h^{\alpha\beta}) = {\cal FP}\, \Box^{-1}_{\rm ret} [{\cal
M}(\Lambda^{\alpha\beta})] - \frac{4G}{c^4} \sum^{+\infty}_{l=0}
\frac{(-)^l}{l!} \partial_L \left\{ \frac{1}{r} {\cal
F}^{\alpha\beta}_L (t-r/c) \right\}\;,
\end{equation}
where the ``multipole moments'' are given by 

\begin{equation}\label{13}
{\cal F}^{\alpha\beta}_L (u) = {\cal FP}\int d^3 {\bf y}~\!  {\hat
 y}_L \, \int^1_{-1} dz ~\!\delta_l(z)~\!\overline{\tau}^{\alpha\beta}
 ({\bf y}, u-z|{\bf y}|/c)\;.
\end{equation} 
Here, ${\overline \tau}^{\alpha\beta}$ denotes the post-Newtonian
expansion of the stress-energy pseudo-tensor defined by Eq. (\ref{2}).
\end{theorem}
The function $\delta_l(z)$ entering the moments is \footnote{This
function approaches the Dirac delta-function in the limit of large
$l$~: $\lim_{~l \to +\infty}\delta_l (z)=\delta (z)$, and its integral
is normalized to one~: $\int^1_{-1} dz~\!\delta_l (z) = 1$.}

\begin{equation}\label{14}
\delta_l (z) = {(2l+1)!!\over 2^{l+1} l!} (1-z^2)^l\;. 
\end{equation} 
A different formalism, alternative to Theorem \ref{th5}, has been
developed by Will and Wiseman \cite{WWi96}, and proved
\cite{Behlers,Bliving} to be strictly equivalent to ours.

With Theorem \ref{th5} in hands, one can obtain the expressions of the
source multipole moments $I_L$ and $J_L$ parametrizing the exterior
field of Sections \ref{II}-\ref{III}. These are the ones to be
inserted into the linearized metric (\ref{3})-(\ref{6}) and iterated
by means of the post-Minkowskian algorithm (\ref{7})-(\ref{8}).

\begin{eqnarray}
I_L(u)&=& {\cal FP} \int d^3{\bf y} \int^1_{-1} dz\biggl\{
 \delta_l\hat y_L\Sigma -\frac{4(2l+1)}{c^2(l+1)(2l+3)} \delta_{l+1}
 \hat y_{iL} \Sigma^{(1)}_i\nonumber\\
 &+&\frac{2(2l+1)}{c^4(l+1)(l+2)(2l+5)} \delta_{l+2} \hat y_{ijL}
 \Sigma^{(2)}_{ij} \biggr\} ({\bf y},u-z |{\bf y}|/c)\;, \label{15}\\
 \nonumber \\ J_L(u)&=& {\cal FP} \int d^3{\bf y}\int^1_{-1}
 dz~\!\varepsilon_{ab<i_l} \biggl\{ \delta_l {\hat y}_{L-1>a} \Sigma_b
 \nonumber\\ &-&\frac{2l+1}{c^2(l+2)(2l+3)} \delta_{l+1} \hat
 y_{L-1>ac} \Sigma^{(1)}_{bc} \biggr\} ({\bf y},u-z |{\bf
 y}|/c)\;,\label{16}
\end{eqnarray}
where we have posed

\begin{equation}\label{17}
\Sigma=\frac{\overline{\tau}^{00}+\overline{\tau}^{ii}}{c^2}\;,\quad\Sigma_i=
\frac{\overline{\tau}^{0i}}{c}\;,\quad\Sigma_{ij}=\overline{\tau}^{ij}\;.
\end{equation}

\section{Post-Newtonian field in the near zone}\label{VI}

The post-Newtonian metric can be systematically computed by means of a
specific variant of the operator of the ``symmetric'' potentials,

\begin{equation}\label{18}
{\cal I}^{-1} = \frac{1}{\Delta-\frac{1}{c^2}\partial_t^2}
= \sum_{k=0}^{+\infty}\frac{1}{c^{2k}}\partial_t^{2k}\Delta^{-k-1}\;.
\end{equation}

\begin{theorem}\label{th6}\cite{PB01}
The expression of the post-Newtonian field inside the source,
satisfying correct boundary conditions at infinity (no incoming
radiation), follows from the matching equation (\ref{11}) as

\begin{eqnarray}
\overline{h}^{\alpha\beta}&=&\frac{16\pi G}{c^4} {\cal FP}~\!{\cal
I}^{-1}\left[\overline{\tau}^{\alpha\beta}\right]\nonumber\\ &-& {4G\over c^4}
\sum^{+\infty}_{l=0} {(-)^l\over l!} \partial_L \left\{
\frac{{\cal A}^{\alpha\beta}_L (t-r/c)-{\cal A}^{\alpha\beta}_L
(t+r/c)}{2r} \right\}\;.\label{19}
\end{eqnarray} 
The ``anti-symmetric'' waves are parametrized by ${\cal
A}^{\alpha\beta}_L(u) ={\cal F}^{\alpha\beta}_L(u) +{\cal
R}^{\alpha\beta}_L(u)$, where ${\cal F}^{\alpha\beta}_L$ is given by
Eq. (\ref{13}), and where \footnote{We denote
$\gamma_l(z)=-2\delta_l(z)$. The normalization of $\gamma_l(z)$~:
$\int^{+\infty}_{1} dz~\!\gamma_l (z) = 1$ (where the integral is
computed by analytic continuation in $l\in \mathbb{C}$), is consistent
with that of $\delta_l(z)$.}

\begin{equation}\label{20}
{\cal R}^{\alpha\beta}_L(u) = {\cal FP}\int d^3{\bf y}~\!{\hat y}_L
\int_1^{+\infty}dz~\!\gamma_l(z)~\!{\cal M}(\tau^{\alpha\beta})({\bf
y},u-z |{\bf y}|/c)\;.
\end{equation}
\end{theorem}
The anti-symmetric waves are associated with radiation-reaction
effects.  For instance the ${\cal F}^{\alpha\beta}_L$-contribution
yields the dominant radiation reaction force at the 2.5PN order
\cite{BuTh70,Bu71}, while the ${\cal R}^{\alpha\beta}_L$-contribution
contains the gravitational-wave tails in the radiation reaction, which
arise at the 4PN order \cite{BD88,B93}.

\bigskip
\centerline{\bf{PART B~: COMPACT BINARIES}}
\bigskip

We recall that the forthcoming detection and analysis of the
gravitational waves emitted by inspiralling compact binaries should
necessitate the prior knowledge of the equations of motion and
gravitational radiation up to the high 3PN order \cite{3mn,CFPS93}.

\section{The problem of self-field regularization}\label{VII}

It makes sense in post-Newtonian approximations to model the compact
objects by point-like particles, but we then face the problem of the
regularization of the infinite self-field of the particles. The
Hadamard regularization has been extensively used in this field, but
it turned out, as we now review, to yield an unknow coefficient at the
3PN order (and not before that order).

Jaranowski and Sch\"afer \cite{JaraS98,JaraS99} noticed that, in their
computation of the equations of motion within the ADM-Hamiltonian
formulation of general relativity, the ``standard'' Hadamard
regularization leads to some ambiguous results for the computation of
certain integrals at the 3PN order. They showed that there are {\it
two and only two} types of ambiguous terms in the 3PN Hamiltonian,
which they parametrized by two unknown numerical coefficients
$\omega_{\rm static}$ and $\omega_{\rm kinetic}$.

Motivated by this result, Blanchet and Faye \cite{BFreg,BFregM}
introduced an ``improved'' Hadamard regularization, based on a theory
of pseudo-functions and generalized distributional derivatives, which
is free of ambiguities, and leads to unique results for the
computation of all the integrals occuring at the 3PN
order. Unfortunately, this regularization turned out to be in a sense
incomplete, because it was found \cite{BF00,BFeom} that the 3PN
equations of motion involve {\it one and only one} unknown numerical
constant, called $\lambda$, which cannot be determined within the
method.  The comparison of this result with the work of Jaranowski and
Sch\"afer \cite{JaraS98,JaraS99}, on the basis of the computation of
the invariant energy of binaries moving on circular orbits, revealed
\cite{BF00} that

\begin{eqnarray}
\omega_{\rm kinetic}&=&\frac{41}{24}\;,\label{21}\\
\omega_{\rm
static}&=&-\frac{11}{3}\lambda-\frac{1987}{840}\;.\label{22}
\end{eqnarray}
The ambiguity $\omega_{\rm kinetic}$ is therefore fixed, while
$\lambda$ is equivalent to the other ambiguity $\omega_{\rm static}$.

Damour, Jaranowski and Sch\"afer \cite{DJS00} recovered the value of
$\omega_{\rm kinetic}$ given in Eq. (\ref{21}) by proving that this
value is the unique one for which the global Poincar\'e invariance of
their formalism is verified. Since the coordinate conditions
associated with the ADM approach do not manifestly respect the
Poincar\'e symmetry, they had to prove that the Hamiltonian is
compatible with the existence of generators for the Poincar\'e
algebra. By contrast, the harmonic-coordinates condition preserves the
Poincar\'e invariance, and therefore the associated equations of
motion should be Lorentz-invariant, as was indeed found to be the case
by Blanchet and Faye \cite{BF00,BFeom}, thanks to their use of a
Lorentz-invariant regularization \cite{BFregM} (hence their
determination of $\omega_{\rm kinetic}$).
 
More recently, the other parameter $\omega_{\rm static}$ was computed
by Damour, Jaranowski and Sch\"afer \cite{DJS01b} by means of a
dimensional regularization, instead of an Hadamard-type one, within
the ADM-Hamiltonian formalism.  The result, which in principle fixes
$\lambda$ according to Eq. (\ref{22}), is

\begin{equation}\label{23}
\omega_{\rm static}=0~~\Longleftrightarrow~~\lambda=-\frac{1987}{3080}\;.
\end{equation}
As Damour {\it et al} \cite{DJS01b} argue, clearing up the ambiguity
is made possible by the fact that the dimensional regularization,
contrary to the Hadamard one, respects all the basic properties of the
algebraic and differential calculus of ordinary functions
\footnote{For instance, the Hadamard regularization unavoidably
violates at some stage the Leibniz rule for the differentiation of a
product.}.

Blanchet, Iyer and Joguet \cite{BIJ01}, in their computation of the
3PN radiation field of two point particles, used the (standard)
Hadamard regularization and found necessary to introduce three
additional regularization constants $\xi$, $\kappa$ and $\zeta$,
playing a role analogous to the equation-of-motion constant
$\lambda$. Such unknowns come from the computation of the 3PN binary's
quadrupole moment $I_{ij}$. It was found \cite{BIJ01} that the total
gravitational-wave flux, in the case of circular orbits, depends in
fact only on a single combination of these three constants,

\begin{equation}\label{24}
\theta = \xi+2\kappa+\zeta\;.
\end{equation}

\section{The equations of binary motion}\label{VIII}

The equations are expressed in Newtonian-like form, by means of the
coordinate positions $y_1^i$, $y_2^i$, and velocities
$v_1^i=dy_1^i/dt$, $v_2^i=dy_2^i/dt$, of the bodies in the harmonic
coordinate system. The 3PN acceleration of one of the bodies, say 1,
reads

\begin{equation}\label{25}
\frac{dv_1^i}{dt} = A_{\rm N}^i + \frac{1}{c^2}A_{\rm 1PN}^i +
\frac{1}{c^4}A_{\rm 2PN}^i + \frac{1}{c^5}A_{\rm 2.5PN}^i +
\frac{1}{c^6}A_{\rm 3PN}^i + {\cal O}\left(\frac{1}{c^7}\right)\;,
\end{equation}
The first term is the famous Newtonian law,

\begin{equation}\label{26}
A_{\rm N}^i = -\frac{G m_2}{r_{12}^2}n_{12}^i\;, 
\end{equation}
(where $r_{12}=|{\bf y}_1-{\bf y}_2|$ and ${\bf n}_{12}=({\bf
y}_1-{\bf y}_2)/r_{12}$). All the other terms represent the successive
post-Newtonian corrections. The 3PN equations of motion, that we show
below, possess the correct perturbative limit given by the geodesics
of the Schwarzschild metric, stay invariant when we perform a global
Lorentz transformation, and are derivable from a Lagrangian or
Hamiltonian formulation (when the radiation-reaction term is
neglected).  The 1PN term is the Einstein-Infeld-Hoffmann \cite{EIH}
acceleration~:

\begin{eqnarray}\label{27}
A_{\rm 1PN}^i &=& \bigg[5\frac{G^2 m_1 m_2}{r_{12}^3} + 4\frac{
G^2 m_2^2}{r_{12}^3 }+ \frac{G m_2}{r_{12}^2} \bigg(\frac{3}{2}
(n_{12}v_2)^2 - v_1^2 + 4 (v_1v_2) - 2 v_2^2\bigg)\bigg] n_{12}^i
\nonumber\\ &+& \frac{G m_2}{r_{12}^2} \bigg[4 (n_{12}v_1) -
3 (n_{12}v_2)\bigg] v_{12}^i\;.
\end{eqnarray}
The parenthesis indicate the usual Euclidean scalar product,
e.g. $(n_{12}v_1)={\bf n}_{12}.{\bf v}_1$, and
$v_{12}^i\equiv v_{1}^i-v_{2}^i$. At the 2PN order, the result is
\cite{DD81a,D83a,Kop85,GKop86,BFP98,IFA01,PW02}

\begin{eqnarray} \label{28}
A_{\rm 2PN}^i &=& \bigg[-\frac{57}{4}\frac{G^3 m_1^2 m_2}{r_{12}^4}-
\frac{69}{2}\frac{G^3 m_1 m_2^2}{r_{12}^4}- 9\frac{G^3
m_2^3}{r_{12}^4} \nonumber\\&+& \frac{G m_2}{r_{12}^2}
\bigg(-\frac{15}{8} (n_{12}v_2)^4 + \frac{3}{2} (n_{12}v_2)^2 v_1^2
\nonumber\\ &-& 6 (n_{12}v_2)^2 (v_1v_2) - 2 (v_1v_2)^2 + \frac{9}{2}
(n_{12}v_2)^2 v_2^2 + 4 (v_1v_2) v_2^2 - 2 v_2^4\bigg) \nonumber\\ &+&
\frac{G^2 m_1 m_2}{ r_{12}^3} \bigg(\frac{39}{2} (n_{12}v_1)^2 - 39
(n_{12}v_1) (n_{12}v_2) \nonumber\\&+& \frac{17}{2} (n_{12}v_2)^2 -
\frac{15}{4} v_1^2 - \frac{5}{2} (v_1v_2) + \frac{5}{4} v_2^2\bigg)
\nonumber\\ &+& \frac{G^2 m_2^2}{r_{12}^3} \bigg(2 (n_{12}v_1)^2 - 4
(n_{12}v_1) (n_{12}v_2) - 6 (n_{12}v_2)^2 - 8 (v_1v_2) + 4
v_2^2\bigg)\bigg] n_{12}^i \nonumber\\ &+& \bigg[\frac{G^2
m_2^2}{r_{12}^3} \bigg(-2 (n_{12}v_1) - 2 (n_{12}v_2)\bigg)
\nonumber\\&+& \frac{G^2 m_1 m_2}{r_{12}^3} \bigg(-\frac{63}{4}
(n_{12}v_1) + \frac{55}{4} (n_{12}v_2)\bigg) \nonumber\\ &+& \frac{G
m_2}{r_{12}^2} \bigg(-6 (n_{12}v_1) (n_{12}v_2)^2 + \frac{9}{2}
(n_{12}v_2)^3 + (n_{12}v_2) v_1^2 - 4 (n_{12}v_1) (v_1v_2) \nonumber\\
&+& 4 (n_{12}v_2) (v_1v_2) + 4 (n_{12}v_1) v_2^2 - 5 (n_{12}v_2)
v_2^2\bigg)\bigg] v_{12}^i\;.
\end{eqnarray}
The 2.5PN term -- an ``odd'' term $\sim 1/c^5$ -- corresponds to the
damping effect of radiation reaction~:

\begin{eqnarray} \label{29}
A_{\rm 2.5PN}^i &=& \frac{4}{5}\frac{G^2 m_1 m_2}{r_{12}^3} \bigg[ -
6\frac{G m_1}{r_{12}} + \frac{52}{3}\frac{G m_2}{r_{12}} + 3v_{12}^2
\bigg] (n_{12}v_{12}) n_{12}^i \nonumber\\ &+& \frac{4}{5}\frac{G^2 m_1
m_2}{r_{12}^3}\bigg[2\frac{G m_1}{r_{12}} - 8\frac{G m_2}{r_{12}} -
v_{12}^2 \bigg] v_{12}^i\;.
\end{eqnarray}
Finally, the very lengthy 3PN coefficient is given by
\cite{BF00,BFeom}

\begin{eqnarray} \label{30}
A_{\rm 3PN}^i &=& \bigg[\frac{G m_2}{r_{12}^2} \bigg(\frac{35}{16}
(n_{12}v_2)^6 - \frac{15}{8} (n_{12}v_2)^4 v_1^2 + \frac{15}{2}
(n_{12}v_2)^4 (v_1v_2) \nonumber\\&+& 3 (n_{12}v_2)^2 (v_1v_2)^2 -
\frac{15}{2} (n_{12}v_2)^4 v_2^2 + \frac{3}{2} (n_{12}v_2)^2 v_1^2
v_2^2 \nonumber\\&-& 12 (n_{12}v_2)^2 (v_1v_2) v_2^2 - 2 (v_1v_2)^2 v_2^2
\nonumber\\ &+& \frac{15}{2} (n_{12}v_2)^2 v_2^4 + 4 (v_1v_2) v_2^4 -
2 v_2^6\bigg) + \frac{G^2 m_1 m_2}{ r_{12}^3 } \bigg(-\frac{171}{8}
(n_{12}v_1)^4 \nonumber\\ &+& \frac{171}{2} (n_{12}v_1)^3 (n_{12}v_2)
- \frac{723}{4} (n_{12}v_1)^2 (n_{12}v_2)^2 + \frac{383}{2}
(n_{12}v_1) (n_{12}v_2)^3 \nonumber\\
&-& \frac{455}{8} (n_{12}v_2)^4
+ \frac{229}{4} (n_{12}v_1)^2 v_1^2 - \frac{205}{2} (n_{12}v_1)
(n_{12}v_2) v_1^2 \nonumber\\&+& \frac{191}{4} (n_{12}v_2)^2 v_1^2 - \frac{91}{8}
v_1^4 -\frac{229}{2} (n_{12}v_1)^2 (v_1v_2) + 244 (n_{12}v_1)
(n_{12}v_2) (v_1v_2) \nonumber\\&-& \frac{225}{2} (n_{12}v_2)^2 (v_1v_2)
+ \frac{91}{2} v_1^2 (v_1v_2) - \frac{177}{4} (v_1v_2)^2 +
\frac{229}{4} (n_{12}v_1)^2 v_2^2 \nonumber\\&-& \frac{283}{2} (n_{12}v_1)
(n_{12}v_2) v_2^2 + \frac{259}{4} (n_{12}v_2)^2 v_2^2 -
\frac{91}{4} v_1^2 v_2^2 + 43 (v_1v_2) v_2^2 - \frac{81}{8}
v_2^4\bigg) \nonumber\\&+& \frac{G^2 m_2^2}{ r_{12}^3 } \bigg(-6 (n_{12}v_1)^2
(n_{12}v_2)^2 + 12 (n_{12}v_1) (n_{12}v_2)^3 + 6
(n_{12}v_2)^4 \nonumber\\&+& 4 (n_{12}v_1) (n_{12}v_2) (v_1v_2) + 12 (n_{12}v_2)^2
(v_1v_2) + 4 (v_1v_2)^2 \nonumber\\&-& 4 (n_{12}v_1) (n_{12}v_2) v_2^2
- 12 (n_{12}v_2)^2 v_2^2 - 8 (v_1v_2) v_2^2 + 4 v_2^4\bigg)
\nonumber\\ &+& \frac{G^3 m_2^3}{r_{12}^4} \bigg(-(n_{12}v_1)^2 + 2
(n_{12}v_1) (n_{12}v_2) + \frac{43}{2} (n_{12}v_2)^2 + 18 (v_1v_2) - 9
v_2^2\bigg) \nonumber\\ &+& \frac{G^3 m_1 m_2^2}{r_{12}^4}
\bigg(\frac{415}{8} (n_{12}v_1)^2 - \frac{375}{4} (n_{12}v_1)
(n_{12}v_2) + \frac{1113}{8} (n_{12}v_2)^2 \nonumber\\ &-&
\frac{615}{64}(n_{12}v_{12})^2 \pi^2 + \frac{123}{64} v_{12}^2\pi^2 +
18 v_1^2 + 33 (v_1v_2) - \frac{33}{2} v_2^2 \bigg) \nonumber\\ &+&
\frac{G^3 m_1^2 m_2}{r_{12}^4} \bigg(-\frac{45887}{168} (n_{12}v_1)^2
+ \frac{24025}{42} (n_{12}v_1) (n_{12}v_2) \nonumber\\&-& \frac{10469}{42}
(n_{12}v_2)^2 + \frac{48197}{840} v_1^2 -
\frac{36227}{420} (v_1v_2) + \frac{36227}{840} v_2^2 \nonumber\\&+& 110
(n_{12}v_{12})^2 \ln \left(\frac{r_{12}}{r'_1} \right) - 22 v_{12}^2
\ln \left(\frac{r_{12}}{r'_1} \right) \bigg) \nonumber\\ &+& \frac{G^4
m_1^3 m_2}{r_{12}^5} \bigg(-\frac{3187}{1260} + \frac{44}{3} \ln
\left(\frac{r_{12}}{r'_1} \right)\bigg) \nonumber\\&+& \frac{G^4 m_1^2
m_2^2}{r_{12}^5} \bigg(\frac{34763}{210} - \frac{44}{3}\lambda -
\frac{41}{16} \pi^2\bigg) +16\frac{G^4 m_2^4}{r_{12}^5} \nonumber\\ &+& \frac{G^4 m_1
m_2^3}{r_{12}^5} \bigg(\frac{10478}{63} - \frac{44}{3}\lambda -
\frac{41}{16} \pi^2 - \frac{44}{3} \ln \left(\frac{r_{12}}{r'_2}
\right)\bigg) \bigg] n_{12}^i \nonumber\\
&+& \bigg[\frac{G m_2}{r_{12}^2} \bigg(\frac{15}{2} (n_{12}v_1)
(n_{12}v_2)^4 - \frac{45}{8} (n_{12}v_2)^5 - \frac{3}{2} (n_{12}v_2)^3
v_1^2 \nonumber\\&+& 6 (n_{12}v_1) (n_{12}v_2)^2 (v_1v_2) - 6
(n_{12}v_2)^3 (v_1v_2) - 2 (n_{12}v_2) (v_1v_2)^2 \nonumber\\&-& 12
(n_{12}v_1)(n_{12}v_2)^2 v_2^2 + 12 (n_{12}v_2)^3 v_2^2 
+ (n_{12}v_2) v_1^2 v_2^2 \nonumber\\&-& 4 (n_{12}v_1) (v_1v_2) v_2^2 + 8
(n_{12}v_2) (v_1v_2) v_2^2 \nonumber\\&+& 4 (n_{12}v_1) v_2^4 - 7 (n_{12}v_2)
v_2^4\bigg) \nonumber\\ &+& \frac{G^2 m_2^2}{r_{12}^3} \bigg(-2
(n_{12}v_1)^2 (n_{12}v_2) + 8 (n_{12}v_1) (n_{12}v_2)^2 + 2
(n_{12}v_2)^3 \nonumber\\ &+& 2 (n_{12}v_1) (v_1v_2) + 4 (n_{12}v_2)
(v_1v_2) - 2 (n_{12}v_1) v_2^2 - 4 (n_{12}v_2) v_2^2\bigg) \nonumber\\
&+& \frac{G^2 m_1 m_2}{r_{12}^3} \bigg(-\frac{243}{4} (n_{12}v_1)^3 +
\frac{565}{4} (n_{12}v_1)^2 (n_{12}v_2) \nonumber\\&-& \frac{269}{4} (n_{12}v_1)
(n_{12}v_2)^2 - \frac{95}{12} (n_{12}v_2)^3 +
\frac{207}{8} (n_{12}v_1) v_1^2 \nonumber\\&-& \frac{137}{8} (n_{12}v_2) v_1^2 - 36
(n_{12}v_1) (v_1v_2) \nonumber\\ &+& \frac{27}{4} (n_{12}v_2)
(v_1v_2)+ \frac{81}{8} (n_{12}v_1) v_2^2 + \frac{83}{8} (n_{12}v_2)
v_2^2\bigg) \nonumber\\&+& \frac{G^3 m_2^3}{r_{12}^4} \bigg(4 (n_{12}v_1) + 5
(n_{12}v_2)\bigg) \nonumber\\ &+& \frac{G^3 m_1 m_2^2}{r_{12}^4}
\bigg(-\frac{307}{8} (n_{12}v_1) + \frac{479}{8} (n_{12}v_2) +
\frac{123}{32} (n_{12}v_{12}) \pi^2 \bigg) \nonumber\\ &+& \frac{G^3
m_1^2 m_2}{r_{12}^4} \bigg(\frac{31397}{420} (n_{12}v_1) -
\frac{36227}{420} (n_{12}v_2) \nonumber\\&-& 44 (n_{12}v_{12}) \ln
\left(\frac{r_{12}}{r'_1} \right) \bigg)\bigg] v_{12}^i \;.
\end{eqnarray}
The latter harmonic-coordinates equations depend at the 3PN order on
two arbitrary constants $r'_1$ and $r'_2$ parametrizing some
logarithmic terms, however these constants are not physical in the
sense that they can be removed by a coordinate transformation. By
contrast, the equations of motion in ADM-coordinates
\cite{JaraS98,JaraS99,DJS00} do not involve logarithms and are free of
such type of constants.

\section{Energy of circular-orbit binaries}\label{IX}

Most inspiralling compact binaries will have been circularized at the
time when they become visible by the LIGO and VIRGO detectors.  In the
case of circular orbits the equations (\ref{25})-(\ref{30}) simplify
drastically. Also we can translate the origin of coordinates to the
binary's center of mass by imposing that the mass-dipole be $I_i = 0$
(notation of Part A).  Mass parameters are the total mass $m=m_1+m_2$,
the reduced mass $\mu=m_1m_2/m$, and the useful symmetric mass ratio

\begin{equation}\label{31}
\nu\equiv \frac{\mu}{m}\equiv \frac{m_1m_2}{(m_1+m_2)^2}\;,
\end{equation}
such that $0<\nu\leq \frac{1}{4}$, with $\nu=\frac{1}{4}$ in the case
of equal masses, and $\nu\to 0$ in the ``test-mass'' limit for one of
the bodies. The energy $E$ of the binary in the center-of-mass frame
is deduced from the 3PN harmonic-coordinates Lagrangian \cite{ABF01}
and expressed in invariant form (the same in different coordinate
systems).  This is achieved with the help of the angular orbital
frequency $\omega$ of the circular orbit, or, more conveniently, the
parameter defined by

\begin{equation}\label{32}
x \equiv \left(\frac{G m \omega}{c^3}\right)^{2/3}\;.
\end{equation}
After replacement we discover that the two gauge constants $r'_1$ and
$r'_2$ cancel out. The end result is

\begin{eqnarray}\label{33}
E &=& -\frac{\mu c^2 x}{2} \biggl\{ 1
 +\left(-\frac{3}{4}-\frac{1}{12}\nu\right) x +
 \left(-\frac{27}{8}+\frac{19}{8}\nu -\frac{1}{24}\nu^2\right) x^2
 \nonumber\\
 &+&\left(-\frac{675}{64}+\left[\frac{209323}{4032}-\frac{205}{96}\pi^2
 -\frac{110}{9}\lambda\right]\nu-\frac{155}{96}\nu^2
 -\frac{35}{5184}\nu^3\right)x^3 \biggr\}\;.\quad
\end{eqnarray}

As an application of Eq. (\ref{33}) let us determine the location of
the minimum of the energy $E$ as a function of the orbital frequency
for circular orbits \cite{B01ico}, say $E_{\rm min}=E(\omega_{\rm
min})$.

\begin{figure}[h]
\centerline{\epsfxsize=20pc
\epsfbox{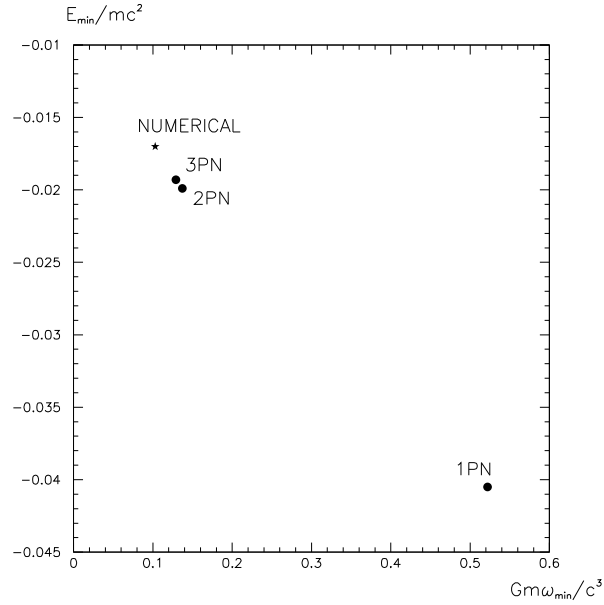}}
\caption{Results for $E_{\rm min}$ in terms of $\omega_{\rm min}$ for
equal-mass binaries ($\nu=\frac{1}{4}$). The asterisk marks the result
calculated by numerical relativity. The point 3PN corresponds to the
value for the ambiguity parameter given by Eq. (\ref{23}).
\label{fig1}}
\end{figure}
\noindent
In Fig. \ref{fig1} we display the results given by the successive 1PN,
2PN and 3PN orders [the 1PN and 2PN orders are defined by the obvious
truncation of Eq. (\ref{33})], and compare them with the prediction
recently calculated numerically by Gourgoulhon, Grandcl\'ement and
Bonazzola \cite{GGB1,GGB2} under the assumption of exactly circular
orbits (helicoidal symmetry). As we see the post-Newtonian and
numerical predictions agree reasonably well. The post-Newtonian
approximation seems to converge well toward the ``exact'' solution,
even at this very relativistic stage where the orbital velocity
reaches $v/c\sim (Gm\omega/c^3)^{1/3}\sim 0.5$. However, the 1PN order
is clearly not precise enough at this stage.

\section{Flux of circular-orbit binaries}\label{X}

The calculation of the flux necessitates, following the formalism
described in Part A, the expressions of the source multipole moments
$I_L$ and $J_L$.  The crucial moment to be calculated is the
mass-quadrupole, which is to be determined with the maximal
post-Newtonian precision. It has been obtained by Blanchet, Iyer and
Joguet \cite{BIJ01}, at the 3PN order in the case of circular orbits
\footnote{The 3PN quadrupole is found to involve three ambiguity
constants $\xi$, $\kappa$ and $\zeta$ [see Eq. (\ref{24})], as well as
some $r_0$ coming from the definition of the finite part process
${\cal FP}$ employed in Part A, and some $r'_0$ which is in fact the
``logarithmic barycenter'' of the constants $r'_1$ and $r'_2$ present
in the equations of motion (\ref{30}).  Both $r_0$ and $r'_0$
disappear from the result at the end, in agreement with general
expectations.}. We compute the time-derivatives of the moments with
the help of the 3PN equations of motion (\ref{25})-(\ref{30}), replace
them into the total energy flux developed at the 3PN order, and add
all the non-linear tail and related effects found up to the 3.5PN
order in Ref. \cite{B98tail}. We must also express the result in terms
of the frequency-dependent parameter (\ref{32}). Finally we get
 
\begin{eqnarray}\label{34}
 {\cal L} &=& \frac{32c^5}{5G}\nu^2 x^5 \biggl\{ 1 +
\left(-\frac{1247}{336}-\frac{35}{12}\nu \right) x + 4\pi
x^{3/2}\nonumber \\ &+&
\left(-\frac{44711}{9072}+\frac{9271}{504}\nu+\frac{65}{18}
\nu^2\right) x^2 +\left(-\frac{8191}{672}-\frac{535}{24}\nu\right)\pi
x^{5/2}\nonumber \\
&+&\left(\frac{6643739519}{69854400}+\frac{16}{3}\pi^2-\frac{1712}{105}C
-\frac{856}{105}\ln (16~\!x) \right.\nonumber\\
&+&\left.\left[-\frac{11497453}{272160}+\frac{41}{48}\pi^2
+\frac{176}{9}\lambda-\frac{88}{3}\theta\right]\nu-\frac{94403}{3024}\nu^2
-\frac{775}{324}\nu^3\right) x^3\nonumber\\ &+&
\left(-\frac{16285}{504}+\frac{176419}{1512}\nu
+\frac{19897}{378}\nu^2\right)\pi x^{7/2} \biggr\}\;.
\end{eqnarray}
We have $\theta = \xi+2\kappa+\zeta$ (and $C\simeq 0.577$ is the Euler
constant). The flux (\ref{34}) agrees perfectly, in the test-mass
limit $\nu\to 0$, with the result of linear black-holes perturbations
calculated by Tagoshi and Sasaki \cite{TSasa94}.

As an application of Eqs. (\ref{34}) and (\ref{33}) one can obtain the orbital phase
$\phi=\int \omega dt$ of an inspiralling compact binary at the 3.5PN
order simply by using the energy balance equation

\begin{equation}\label{35}
\frac{dE}{dt}=-{\cal L}~~\Longrightarrow~~\phi = -\int\frac{\omega
dE}{{\cal L}}\;.
\end{equation}
The orbital phase is the crucial observable to be implemented in the
construction of accurate templates for the detection and analysis of
inspiralling-binary signals in the LIGO and VIRGO experiments.


\end{document}